\documentclass[twocolumn]{aastex631}

\shorttitle{The NSC and NSD: Different stellar population and star formation history}
\shortauthors{Nogueras-Lara et al.}
\graphicspath{}
\begin{document}

\title{The nuclear star cluster and nuclear stellar disc of the Milky Way:\\ Different stellar population and star formation history.}

\correspondingauthor{Francisco Nogueras-Lara}
\email{nogueras@mpia.de}

\author[0000-0002-6379-7593]{Francisco Nogueras-Lara}
\affiliation{Max-Planck Institute for Astronomy\\
K\"onigstuhl 17\\
69117 Heidelberg, Germany}

\author{Rainer Sch\"odel}
\affiliation{Instituto de Astrof\'isica de Andaluc\'ia (CSIC)\\
Glorieta de la Astronom\'ia s/n \\
18008 Granada, Spain}

\author{Nadine Neumayer}
\affiliation{Max-Planck Institute for Astronomy\\
K\"onigstuhl 17\\
69117 Heidelberg, Germany}

%% Note that the \and command from previous versions of AASTeX is now
%% depreciated in this version as it is no longer necessary. AASTeX 
%% automatically takes care of all commas and "and"s between authors names.

%% AASTeX 6.31 has the new \collaboration and \nocollaboration commands to
%% provide the collaboration status of a group of authors. These commands 
%% can be used either before or after the list of corresponding authors. The
%% argument for \collaboration is the collaboration identifier. Authors are
%% encouraged to surround collaboration identifiers with ()s. The 
%% \nocollaboration command takes no argument and exists to indicate that
%% the nearby authors are not part of surrounding collaborations.

%% Mark off the abstract in the ``abstract'' environment. 
\begin{abstract}

The Milky Way's nuclear stellar disc (NSD) and nuclear star cluster (NSC) are the main features of the Galactic centre. Nevertheless, their observation is hampered by the extreme source crowding and high extinction. Hence, their relation and formation scenario are not fully clear yet. We aim at detecting the stellar populations from the NSC and the NSD along the line-of-sight towards the NSC, and assess whether they have different stellar populations and star formation histories. We analysed the colour-magnitude diagram, $K_s$ vs. $H-K_s$, of a region of $8.2'\times2.8'$ centred on the NSC, and detected two different stellar groups with different extinctions. We studied their red clumps to find the features associated to each of the stellar populations. We obtained that the two groups of stars correspond to the NSD and the NSC, and found that they have significantly different stellar populations and star formation histories. We detected a double red clump for the NSD population, in agreement with previous work, whereas the NSC presents a more complex structure well fitted by three Gaussian features. We created extinction maps to analyse the extinction variation between the detected stellar groups. We found that the high-extinction layer varies on smaller scales (arc-seconds), and that there is a difference of $A_{K_s}\sim0.6$\,mag between both extinction layers. Finally, we obtained that the distance towards each of the stellar populations is compatible with the Galactic centre distance, and found some evidence of a slightly closer distance for the NSD stars ($\sim360\pm200$\,pc).

\end{abstract}

%% Keywords should appear after the \end{abstract} command. 
%% The AAS Journals now uses Unified Astronomy Thesaurus concepts:
%% https://astrothesaurus.org
%% You will be asked to selected these concepts during the submission process
%% but this old "keyword" functionality is maintained in case authors want
%% to include these concepts in their preprints.
\keywords{Galactic center(565) --- Galaxy structure(622) --- Red giant clump(1370) --- Interstellar dust extinction(837)}

%% From the front matter, we move on to the body of the paper.
%% Sections are demarcated by \section and \subsection, respectively.
%% Observe the use of the LaTeX \label
%% command after the \subsection to give a symbolic KEY to the
%% subsection for cross-referencing in a \ref command.
%% You can use LaTeX's \ref and \label commands to keep track of
%% cross-references to sections, equations, tables, and figures.
%% That way, if you change the order of any elements, LaTeX will
%% automatically renumber them.
%%
%% We recommend that authors also use the natbib \citep
%% and \citet commands to identify citations.  The citations are
%% tied to the reference list via symbolic KEYs. The KEY corresponds
%% to the KEY in the \bibitem in the reference list below. 

\section{Introduction}

The centre of the Milky Way is the closest galactic nucleus located at only $\sim8$\,kpc from the Earth  \citep[e.g.][]{Gravity-Collaboration:2018aa,Do:2019aa}. It allows us to resolve individual stars down to milli-parsec scales and analyse in great detail its structure and stellar population \citep[e.g.][]{Genzel:2010fk,Schodel:2014bn}. Therefore, it is a perfect laboratory to study galactic nuclei and their role in a wider context of galaxy evolution.

The Galactic centre (GC) hosts two main structures: (1) the nuclear stellar disc (NSD), a disc-like structure with a mass of $\sim 7\times10^8$\,M$_\odot$, and radius $\sim 200$\,pc \citep[e.g.][]{Launhardt:2002nx,Nishiyama:2013uq,gallego-cano2019,Sormani:2020aa}, that partially overlaps with the dense clouds of gas of the central molecular zone \citep[e.g.][]{Morris:1996vn}; and (2) The nuclear star cluster (NSC), a massive stellar cluster \citep[$\sim2.5\times10^7$\,M$_\odot$, e.g.][]{Launhardt:2002nx,Schodel:2014bn,Feldmeier-Krause:2017kq,Do:2019aa,Sormani:2020aa} centred on the supermassive black hole Sagittarius\,A*, with an effective radius of $\sim5$\,pc \citep[e.g.][]{Graham:2009lh,Schodel:2011ab,Feldmeier-Krause:2017kq,gallego-cano2019}.

In spite of its proximity, the observation of the GC is hampered by the extreme source crowding and the high interstellar extinction that limits its analysis to the infrared regime \citep[e.g.][]{Nishiyama:2008qa,Schodel:2010fk,Nogueras-Lara:2018aa,Nogueras-Lara:2020aa}. In this way, the relation between the NSC and the NSD, and their formation processes are not well understood yet \citep[e.g.][]{Launhardt:2002nx,Nogueras-Lara:2019ad,Schodel:2020aa}. There is some evidence that the NSC and the NSD host different stellar populations with different star formation histories \citep[SFHs, e.g.][]{Nogueras-Lara:2019ad,Schodel:2020aa,Schultheis:2021wf}. Both components seem to have a predominantly old stellar population ($\gtrsim8$\,Gyr), followed by several billion years of quiescence. There is some evidence for a $\sim3$\,Gyr year old intermediate age population in the NSC, which cannot be found in the NSD. On the other hand, there is evidence of an important $\sim1$\,Gyr star formation event associated to the NSD, that is not found when analysing the stellar population from the NSC \citep{Nogueras-Lara:2019ad,Schodel:2020aa}.

In this paper, we aim to assess whether the stellar populations of the NSC and the NSD are actually different. For this, we analyse the red clump (RC) features \citep[e.g.][]{Girardi:2016fk}, of a region centred on the NSC. We use photometric data in the $H$ and $K_s$ bands and clearly detect two stellar groups with different extinctions that correspond to the NSC and the NSD. We find that the stellar populations belonging to each of the extinction groups are significantly different and explain the differences with different formation histories.

\section{Data}

\subsection{HAWK-I data}

We used $H$ and $K_s$ photometry obtained with the NIR camera High Acuity Wide Field K-band Imager \citep[HAWKI, ][]{Kissler-Patig:2008fr}, placed at the ESO Very Large Telescope in Chile. The observations were taken in 2013, and constituted a pilot study for the GALACTICNUCLEUS (GNS) survey \citep[a high-angular resolution photometric survey of the GC,][]{Nogueras-Lara:2018aa,Nogueras-Lara:2019aa}, and were presented and described in \citet{Nogueras-Lara:2018aa}. We used these data because the observing conditions were better than for the analogous region in the GNS survey \citep[see Table\,1 in][]{Nogueras-Lara:2018aa}. The best conditions are crucial in this most crowded field of the Galaxy and implies that the $H$ and $K_s$ photometry of the 2013 data is around 1\,mag deeper than in the GNS survey.

\begin{figure*}[!htbp]
   \includegraphics[width=\linewidth]{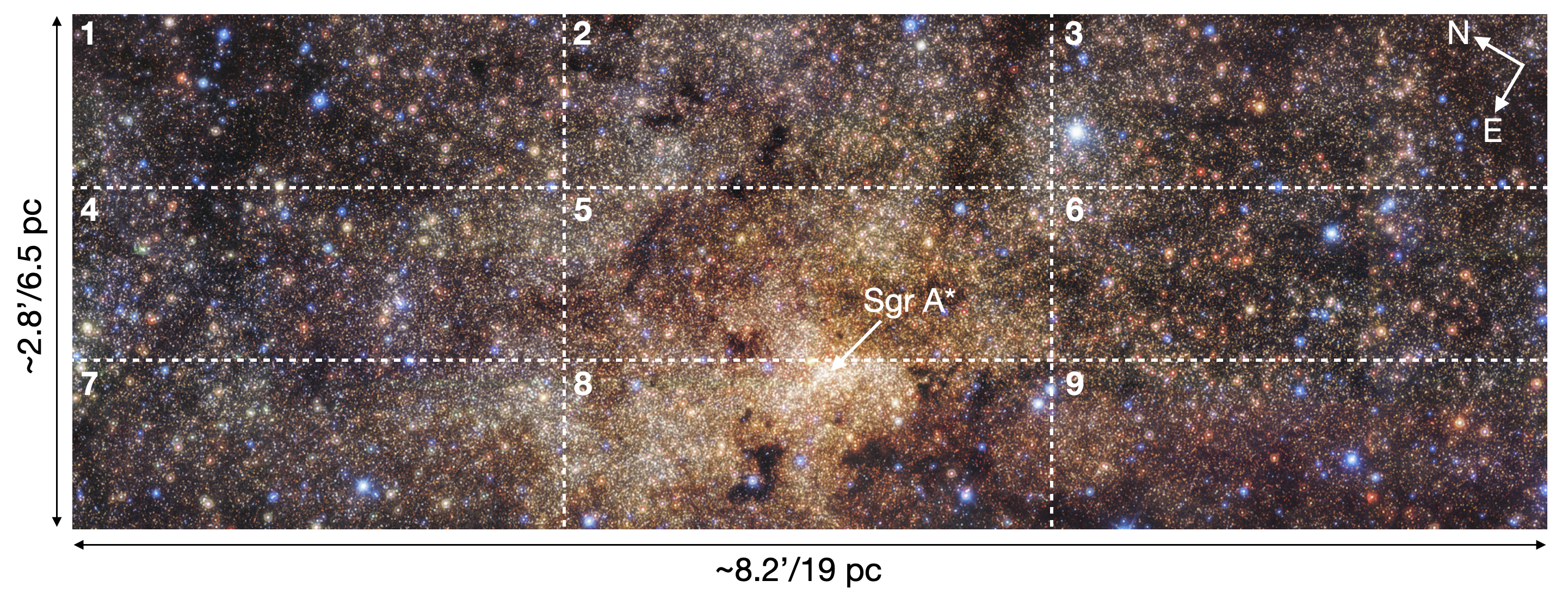}
   \caption{GALACTICNUCLEUS false colour image (using $J$, $H$, and $K_s$ as blue, green, and red, respectively) of the analysed region. The white arrow indicates the position of the supermassive black hole (Sgr\,A*) at the centre of the Galaxy. The white dashed lines indicates the sub-regions used for the analysis shown in Sect.\,\ref{different_los}.}

   \label{scheme}
    \end{figure*}

The observations consist of four independent fixed pointings designed to cover the cross-shaped gap between the four HAWK-I detectors. The observed region was centred on the coordinates (17$^h$45$^m$37.70$^s$,  $-29^\circ$00$'$05.70$''$) covering a total size of $8.2'\times2.8'$ ($\sim19$\,pc $\times$ 6.5\,pc), as shown in Fig.\,\ref{scheme}. Each of the pointings included 512 frames with a detector integration time (DIT) of 0.83\,s \citep[for further details see Sect. 2 in][]{Nogueras-Lara:2018aa}. We processed them following a standard reduction (dark, bias, flat-fielding) and dedicated sky subtraction, to later apply the speckle holography algorithm optimised for crowded fields, as explained in \citet{Schodel:2013fk}. The detectors were treated independently throughout the process. We obtained point spread function photometry for each of them using the {\it Starfinder} algorithm \citep{Diolaiti:2000fk}. We created the final catalogue merging all the detectors, as explained in \citet{Nogueras-Lara:2018aa}. We reached 5$\sigma$ detections at $H\sim21$\,mag and $K_s\sim20$\,mag. The uncertainties were below 0.05\,mag at $H\sim18$\,mag and $K_s\sim17$\,mag. The zero point was calibrated using the SIRIUS/IRSF survey \citep[e.g.][]{Nagayama:2003fk,Nishiyama:2006tx} and its systematic uncertainty was $\sim0.04$\,mag \citep{Nogueras-Lara:2018aa}.

We also used $H$ and $K_s$ data from the GNS survey to analyse a control region to test the obtained results. The survey used a slightly different set up in comparison to the pilot study \citep[for further details see][]{Nogueras-Lara:2018aa,Nogueras-Lara:2019aa}. Both GNS and the pilot study reached $\sim0.2''$ angular resolution in the $JHK_s$ bands.

%The GNS survey also used the speckle holography technique to get a high angular resolution 

%It obtained accurate photometry for more than three million stars with uncertainties below 0.05\,mag at $J\sim21$\,mag, $H\sim19$\,mag, and $K_s\sim 18$\,mag. These are average values considering the whole GNS survey. The uncertainties for the central pointing corresponding to the 2013 data used in this paper, have larger uncertainties \citep[for further details see][]{Nogueras-Lara:2018aa}. The zero point was also calibrated using the SIRIUS/IRSF survey and its systematic uncertainty was $\sim0.04$\,mag \citep{Nogueras-Lara:2018aa}.

\section{Colour-magnitude diagram}

\begin{figure}
   \begin{center}
   \includegraphics[width=\linewidth]{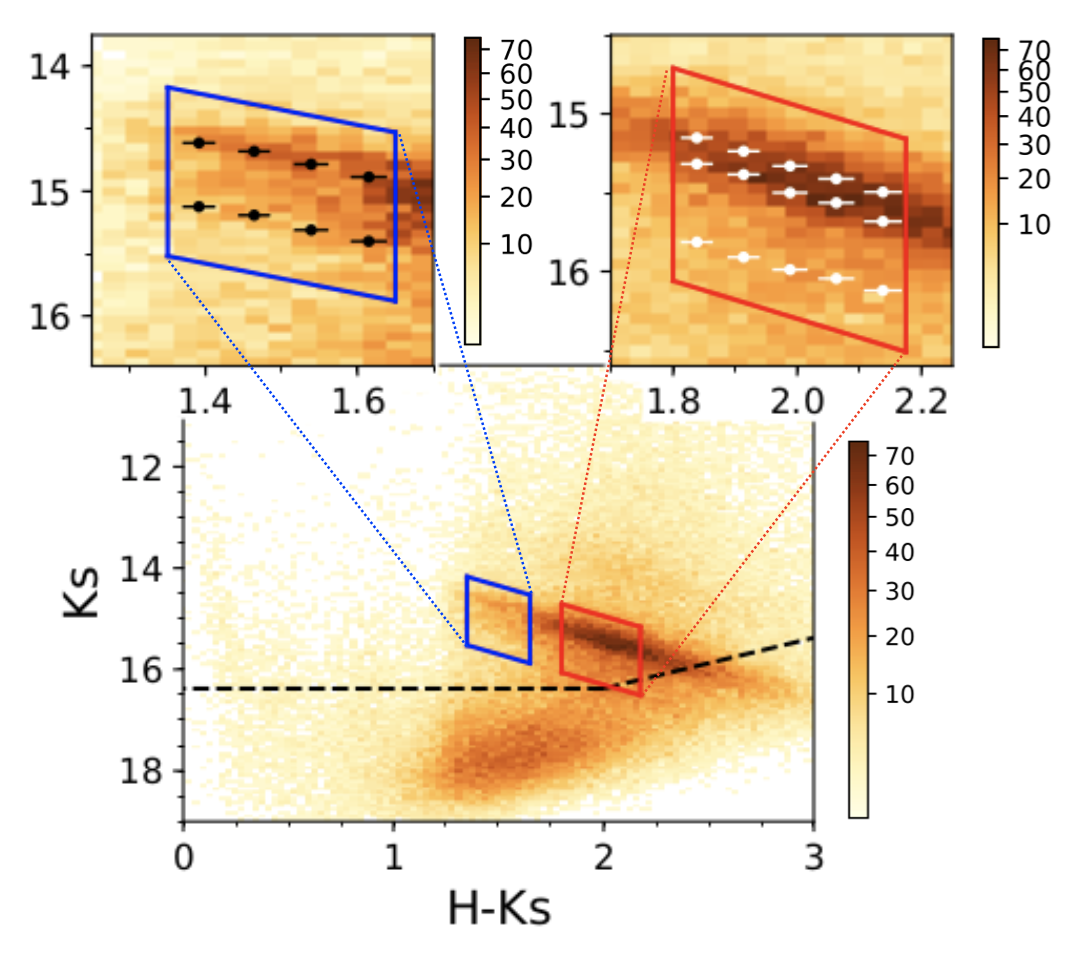}
   \caption{Colour-magnitude diagram $K_s$ vs. $H-K_s$ corresponding to the target region. The colour code corresponds to stellar densities using a power stretch scale. The black dashed line indicates the 2/3 completeness level of the data. The blue and red parallelograms show the reference stars that we consider in Sect.\,\ref{dsp} for the low- and the high-extinction groups. The zoom-in panels depict the best fit found in the RC region using a GMM analysis as specified in Fig.\,\ref{vertical_bins}.}

   \label{CMD}
\end{center}   
    \end{figure}

Figure\,\ref{CMD} shows the colour-magnitude diagram (CMD) $K_s$ vs. $H-K_s$ of the target region. Given the extreme extinction along the line-of-sight towards the GC, a simple colour cut $H-K_s>1.35$\,mag, allowed us to effectively remove the foreground population belonging to the Galactic disc and to the inner Galactic bulge \citep[e.g.][]{Nogueras-Lara:2018ab,Nogueras-Lara:2019aa,Sormani:2020aa,Schultheis:2021wf,Nogueras-Lara:2021uz}. Possible remaining contamination from the inner bulge accounts for less than 25\,\% of the stars beyond that colour cut, according to the analysis carried out by \citet{Schultheis:2021wf} for a similar colour cut and line-of-sight (see their Sect.\,3.1). 

\subsection{Different stellar groups}
\label{dsp}

We visually identified two different stellar groups, that correspond to a low- and a high-extinguished stellar population around $H-K_s\sim1.5$\,mag and $H-K_s\sim2$\,mag, respectively. They appear as clearly distinct populations in an HST/WFC3 CMD F153M vs. F105W-F153M for the innermost $2'\times2'$, as shown in Fig.\,16 in \citet{Nogueras-Lara:2018aa}.

To study the detected stellar populations, we defined two regions in the CMD covering the RC features of each of the stellar populations, as indicated in Fig.\,\ref{CMD}. We selected similar vertical sizes of the selection boxes associated to each of the stellar populations in order to easily compare them in the subsequent analysis. Given that both stellar populations are separated due to the differential extinction, we avoided the stars around $H-K_s\sim1.7$\,mag when defining the regions to be analysed. This is because that region corresponds to a transition region where stars from both stellar populations might appear, complicating the independent analysis. 

We also limited the selection of the faint end of the RC features due to the completeness of the data. Given the extremely high number of stars in the analysed region, the incompleteness is dominated by the source crowding (so the sensitivity is not significantly affecting the completeness in these crowded regions). In this way, we obtained the completeness solution using the technique described by \citet{Eisenhauer:1998tg} and previously applied to similar data sets as the GNS survey \citep[see Sect.\,3.1.2 in][]{Nogueras-Lara:2020aa}. It computes the critical distance at which a star of a given magnitude can be detected near a brighter star, and uses this parameter to estimate the completeness for the real sources in the data set. We applied this method to compute the 2/3 completeness of the data, dividing the observed region into nine smaller subregions (see Fig.\,\ref{scheme}). The final completeness solution was obtained averaging over the results for each of the defined subregions. The reference stars were selected restricting the selection boxes in agreement with the obtained completeness (coloured boxes in Fig.\,\ref{CMD}).

\begin{figure*}
   \includegraphics[width=\linewidth]{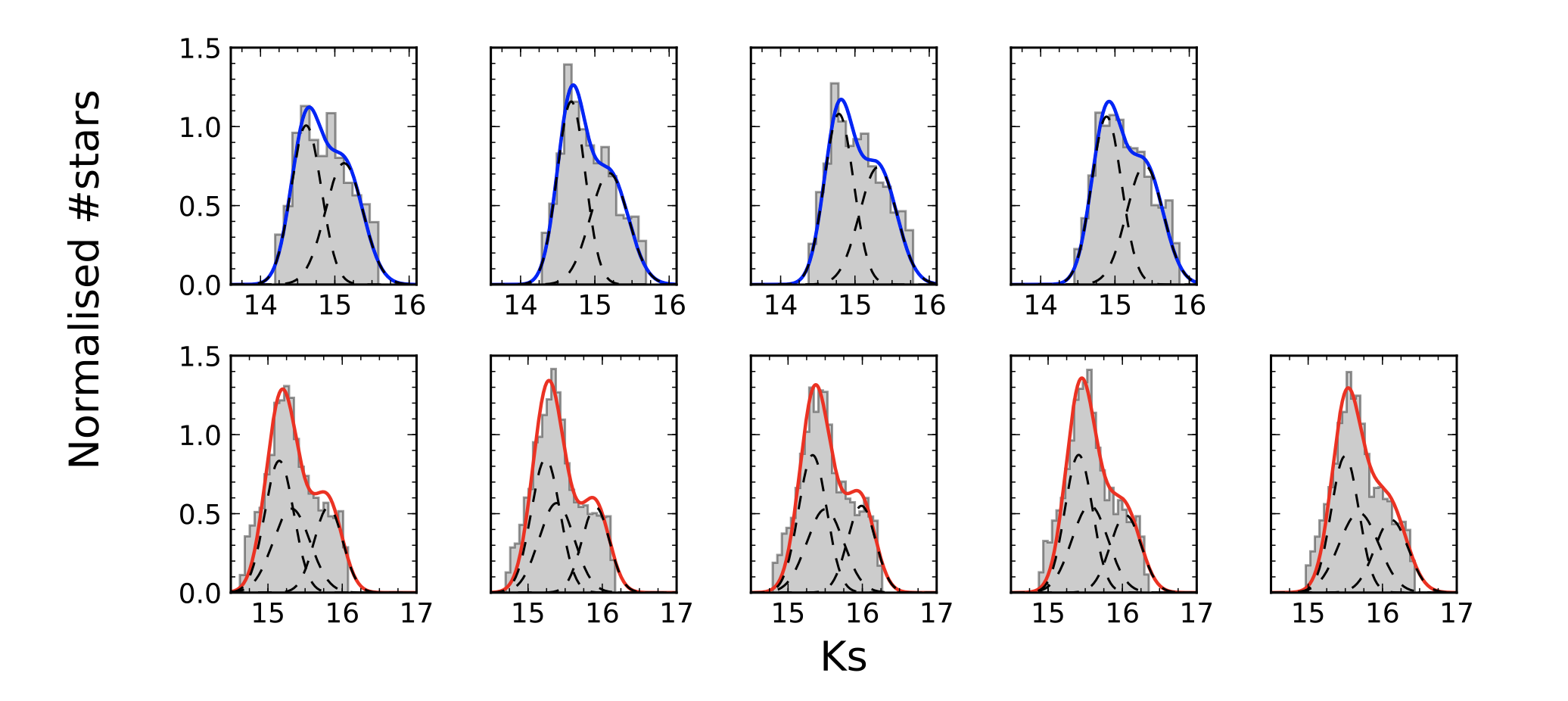}
   \caption{Best-fit models obtained for each of the vertical vertical cuts through the CMD defined in Sect.\ref{dsp}, to analyse the RC features in the CMD $K_s$ vs. $H-K_s$. The grey histograms show the $K_s$ distribution of each of the vertical cuts. The blue and the red solid lines depict the models corresponding to the best fits for the low- and the high- extinguished stellar populations, respectively. The black dashed lines indicate the Gaussian models used for the fits. Upper panels: from left to right, each of the four bins of 0.075\,mag starting from $H-K_s=1.35$\,mag corresponding to the low-extinguished stellar population. Lower panels:  from left to right, each of the five sections of 0.075\,mag starting from $H-K_s=1.8$\,mag corresponding to the high-extinguished stellar population.}

   \label{vertical_bins}
    \end{figure*}

We used the SCIKIT-LEARN python function GaussianMixture \citep[GMM, ][]{Pedregosa:2011aa}, to independently analyse each of the different extinction groups. We divided the selected CMD regions along the $x$-axis ($H-K_s$ colour) into sections of 0.075\,mag width, and used the Bayesian information criterion \citep[BIC, ][]{Schwarz:1978aa} and the Akaike information criterion \citep[AIC][]{Akaike:1974aa}, to compare three different models considering one, two, and three Gaussians distributions to fit the data within each bin. Figure\,\ref{vertical_bins} shows the best fits obtained for each of the vertical bins. In all the cases, we detected a two-Gaussians distribution for the low-extinguished stellar population, whereas the high-extinguished one presented a three-Gaussians distribution. Figure\,\ref{CMD} depicts the obtained results in the zoom-in panel, where the obtained features follow the reddening vector. In particular, we computed the slopes of each of the detected features and obtained: $s_{low\ bright} = 1.24\pm0.13\pm0.12$, $s_{low\ faint} = 1.29\pm0.10\pm0.08$, $s_{high\ bright} = 1.16\pm0.01\pm0.03$, $s_{high\ intermediate} = 1.21\pm0.09\pm0.02$, and $s_{high\ faint} = 1.01\pm0.06\pm0.10$, where the subindices $low$ and $high$ indicate the corresponding extinction group. The uncertainties refer to the statistical and the systematic ones, respectively. The statistical uncertainty was computed using a jackknife resampling method, systematically leaving out one of the data points to calculate the slope \citep[e.g.][]{Nogueras-Lara:2020aa,Nogueras-Lara:2021uq}. The systematics were estimated varying the selection boxes and the size and number of vertical bins. The obtained slopes are consistent within the uncertainties with the slope, $s=1.19\pm0.04$, of the extinction curve derived in \citet{Nogueras-Lara:2019ac,Nogueras-Lara:2020aa} using the GNS catalogue.

\subsection{Comparison with a control region}

\begin{figure}
      \begin{center}
   \includegraphics[width=\linewidth]{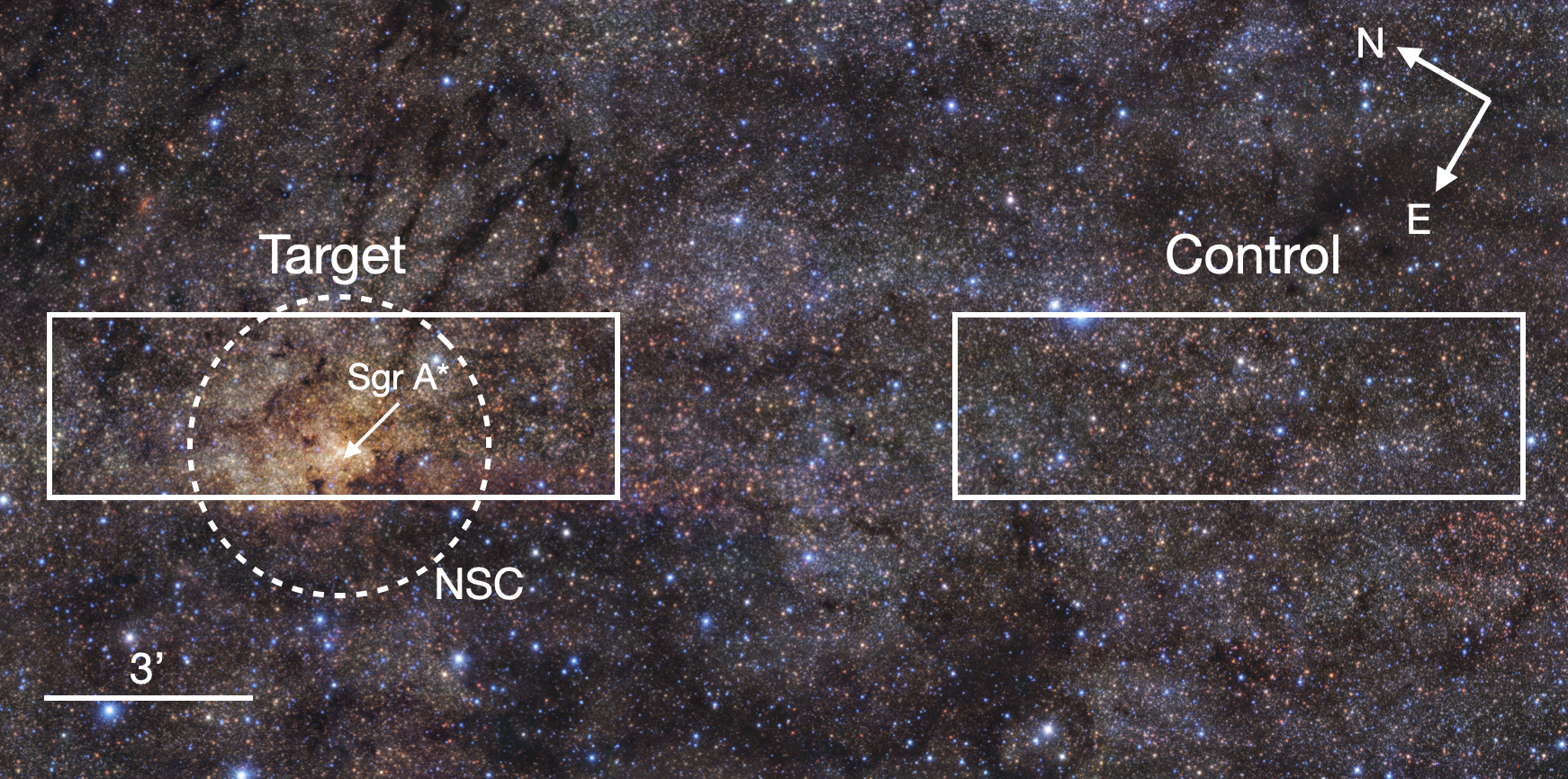}
   \caption{Scheme of the target and control regions selected. The background image corresponds to a false colour image ($JHK_s$) from the GNS survey. The white dashed circle indicates the effective radius of the NSC. The position of Sgr\,A* is indicated as a reference.}

   \label{control_scheme}
   \end{center}
    \end{figure}

We repeated the analysis using a control region of the same size that is located at a distance to the centre of the NSC of more than 6 effective NSC radii \citep[see Fig.\,\ref{control_scheme}, e.g.][]{Schodel:2014bn,Feldmeier-Krause:2017kq,gallego-cano2019}.

\begin{figure}
      \begin{center}
   \includegraphics[width=\linewidth]{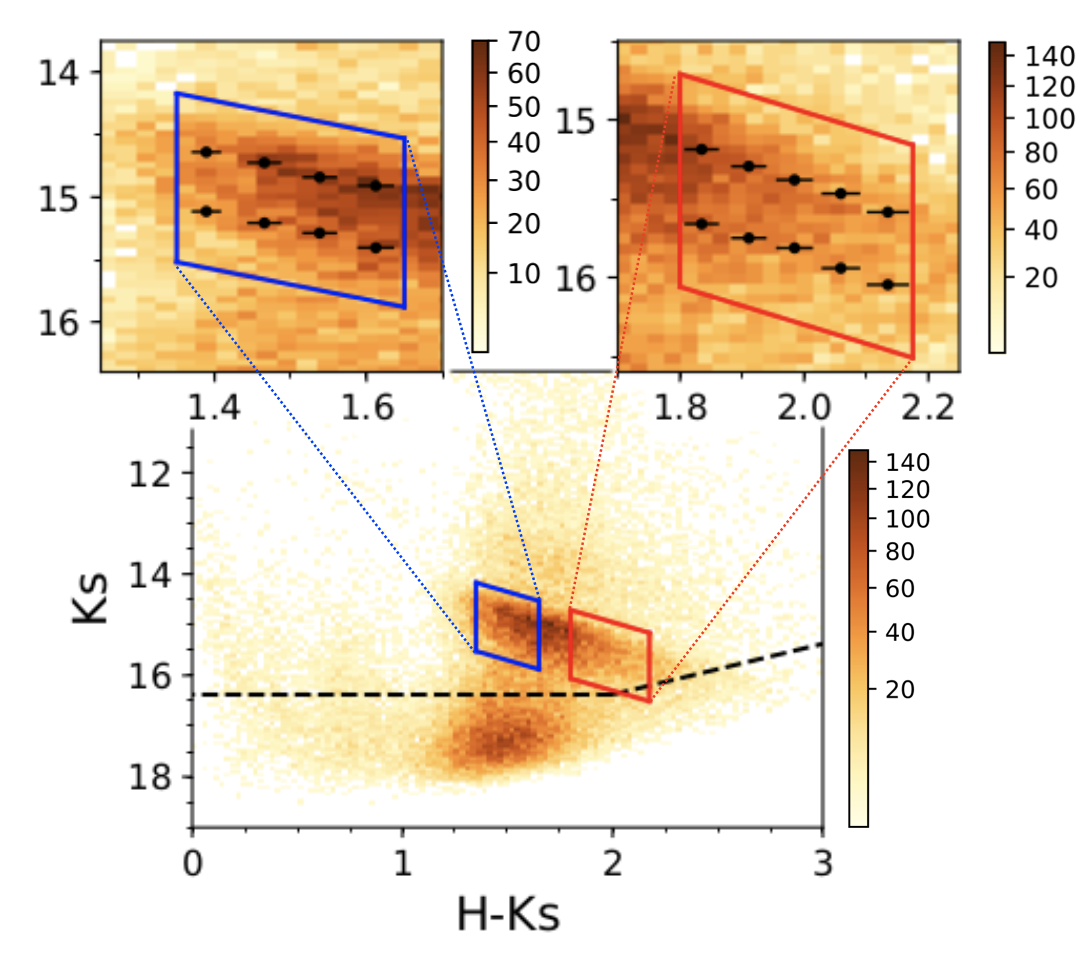}
   \caption{Colour-magnitude diagram $K_s$ vs. $H-K_s$ corresponding to the control region. The colour code corresponds to stellar densities using a power stretch scale. The black dashed line indicates the 2/3 completeness level of the data. The blue and red parallelograms show the reference stars that we consider in Sect.\,\ref{dsp} for the low- and the high-extinction groups. The zoom-in panels depict the best fit found in the RC region using a GMM analysis.}

   \label{CMD_test}
   \end{center}
    \end{figure}

We built a CMD $K_s$ vs. $H-K_s$ (Fig.\,\ref{CMD_test}), and computed the 2/3 completeness due to crowding, using the previously explained technique and the results for the GNS survey in \citet{Nogueras-Lara:2020aa}. We obtained a better data completeness in comparison to the target region because the control region does not contain the NSC, and the crowding is thus less important.

The RC in the control field appears as a continuous feature, without any visible change as a function of extinction. We proceeded in our analysis of the RC region of the CMD exactly as described previously. Figure\,\ref{CMD_test} shows the results. At both high and low extinction, we find the same feature in the RC region of the CMD of the control field. In fact, the high extinction features in the control field can be simply obtained by additional reddening of the low-extinction features. This is markedly different from the target field, where the high-extinction CMD region shows an additional feature associated with the RC. The difference between target and control regions is the presence of the NSC in the former. The NSC stars occupy the high extinction part in the target region CMD.

\subsection{Discussion}
\label{discu}

The presence of several RC features can be explained by two different scenarios: (a) Stellar populations at different distances and (b) RC stars with different ages and/or metallicities \citep[e.g.][]{Ferraro:2009aa,Girardi:2016fk}. To further analyse this, we computed the $K_s$ distance between the detected features in the target region. We obtained:

1) $\Delta K_s\ _{low} = 0.52\pm0.01\pm0.02$\,mag, where the subindex $low$ refers to the low-extinguished stellar population; 2) $\Delta K_s\ _{high\ 1} = 0.16\pm0.01\pm0.01$\,mag, where the subindices $high$ and 1 indicate the difference in magnitude between the brightest and the secondary feature of the high-extinguished stellar population; and 3) $\Delta K_s\ _{high\ 2} = 0.65\pm0.02\pm0.05$\,mag, where the subindex 2 refers to the $K_s$ magnitude distance between the brightest and the faintest features in the high-extinguished stellar population. The $\Delta K_s$ values were obtained averaging over the magnitude differences of the points shown in Fig.\,\ref{CMD}. The first uncertainty corresponds to the statistical one and was estimated via the standard deviation of the measurements. The second one refers to the systematic uncertainty and was computed repeating the analysis varying the limits of the selection box and using different bin-width sizes.

Given the high crowding and extreme extinction in the GC \citep[e.g.][]{Nishiyama:2008qa,Schodel:2010fk,Nogueras-Lara:2018aa,Nogueras-Lara:2020aa}, the scenario (a) (having stellar populations at different distances) is quite unlikely and would be only possible for small distances between the stellar populations. Nevertheless, even for the magnitude difference between  
the brightest and the secondary RC features measured for the high-extinguished stellar group ($\Delta K_s\ _{high\ 1} = 0.16\pm0.01\pm0.01$\,mag), this distance difference would be $\sim 600$\,pc beyond the GC, at the GC distance of $\sim8$\,kpc \citep[e.g.][]{Gravity-Collaboration:2018aa,Do:2019aa}. This would imply that there is a stellar substructure as dense as the NSC located behind the GC, which is highly improbable. The situation becomes even more extreme for the stellar populations associated with larger magnitude differences. Moreover, due to the extreme extinction, stellar populations located at significantly different distances should show some difference in reddening, implying a shift towards redder colours in the $H-K_s$ axis in the CMD with respect to closer components. This was not found for the RC components measured within each of the extinction groups. Thus, we can safely exclude this scenario. Hence, the more likely explanation for the different number of features between the low- and the high-extinguished groups of stars is a difference in their RC stellar population.

The two-Gaussians distribution found for the low-extinguished group of stars is in agreement with the stellar population obtained for the NSD, where previous work found a double RC feature with a magnitude difference of $\Delta K_s \sim 0.5$\,mag \citep[e.g.][]{Rui:2019aa,Nogueras-Lara:2019ad,Nogueras-Lara:2021uq}. This is also in agreement with the analysis of the control region, where we did not observe any difference between the two selection boxes analysed in the RC. We obtained a $\Delta K_s\ _{low} = 0.47\pm0.02\pm0.02$\,mag for the first selection box that corresponds to the low-extinguished stellar group in the target region, and $\Delta K_s\ _{high} = 0.46\pm0.02\pm0.05$\,mag, for the selection box corresponding to the high-extinguished stellar group in the target region. The uncertainties were obtained as previously explained for the target region.

Therefore, we conclude that the low-extinguished stellar group corresponds to the NSD, as it is seen in front of the NSC due to the different extinctions. On the other hand, we believe that the three-Gaussians distribution is associated to the more complex stellar population from the NSC. This is in agreement with previous studies that  suggest that the SFH of the NSC might be different from that of the NSD \citep[e.g.][]{Schodel:2020aa,Schultheis:2021wf}. In particular, according to the SFH derived in \citet{Schodel:2020aa}, the bulk of the stars is old (80\,\% of the stellar mass older than 10\,Gyr), followed by a quiescent period that ended around 3\,Gyr ago with the formation of $\sim 15\,\%$ of the stellar mass. Finally, a few percent of the stars were formed in the past few 100\,Myr. In this way, the three Gaussian features that we detect in our analysis agree with the fit of theoretical $K_s$ luminosity functions in Fig.\,10 of \citet{Schodel:2020aa}, where the two brightest peaks would be produced by the old and intermediate-age stellar populations, and the third one would be the consequence of the formation of the red giant branch bump \citep[RGBB, e.g.][]{Nataf:2011aa,Wegg:2013kx}. The RGBB is also present in the NSD stellar population but partially overlaps with the RC feature from the $\sim$1\,Gyr stellar population, producing a closer secondary peak (see Sect.\,\ref{simul}).

\subsection{Synthetic models}

\subsubsection{CMD simulation}
\label{simul}

\begin{figure}[htp]
      \begin{center}
   \includegraphics[width=\linewidth]{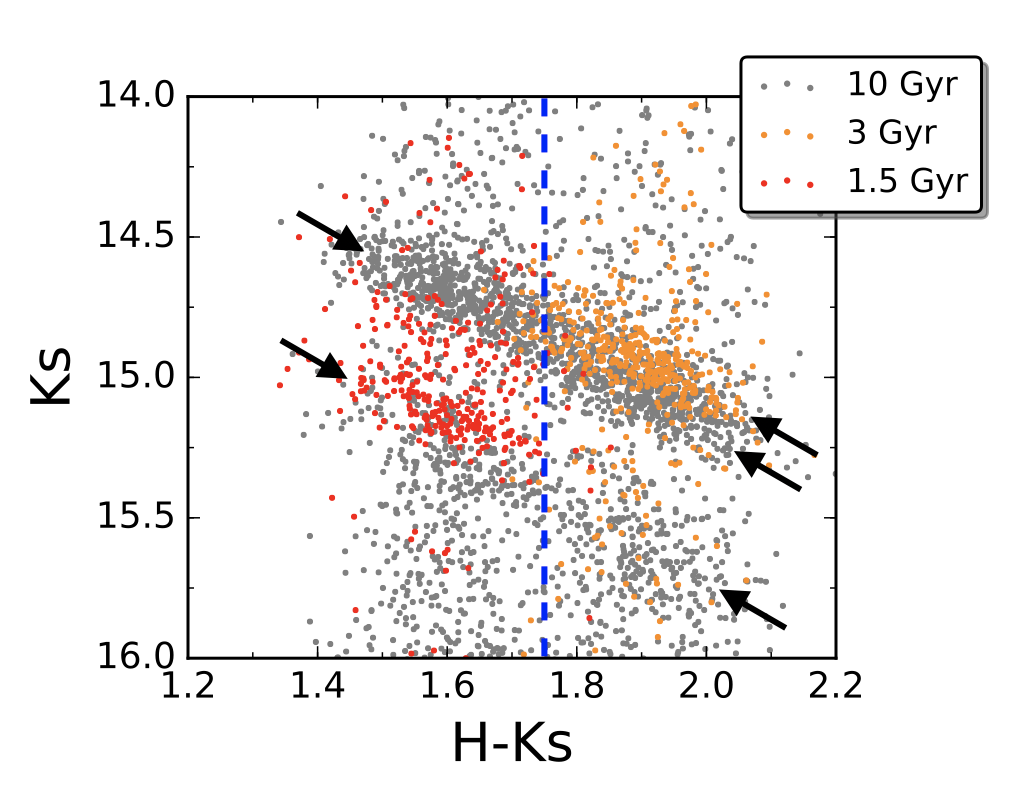}
   \caption{Synthetic CMD $K_s$ vs. $H-K_s$ using Parsec models to simulate the RC distribution found according to the different stellar populations from the NSD and the NSC. The blue dashed line indicates the rough separation between both components due to reddening. The black arrows show the different RC features detected in our analysis.}

   \label{synthetic}
      \end{center}
    \end{figure}

To assess our previous conclusions, we built a simple synthetic CMD $K_s$ vs. $H-K_s$ to simulate the main stellar populations in the NSD and the NSC, using PARSEC models \citep[release v1.2S + COLIBRI S\_37, ][]{Bressan:2012aa,Chen:2014aa,Chen:2015aa,Tang:2014rm,Marigo:2017aa,Pastorelli:2019aa,Pastorelli:2020wz}. Because our purpose is just to detect the main features in the RC region of the CMD, we simply assumed approximately similar masses for the stellar populations belonging to the NSD and the NSC. We only used the stellar models that significantly account for the stars in the RC feature following the results obtained by \citet{Nogueras-Lara:2019ad,Schodel:2020aa}. Namely, for the NSD, we chose an old stellar model of 10\,Gyr ($\sim95\,\%$ of the total stellar mass) and a younger one of 1.5\,Gyr ($\sim5\,\%$ of the total stellar mass). We simulated the NSC using the same old stellar population (10\,Gyr), and an intermediate-age one of 3\,Gyr (accounting for $\sim20\,\%$ of the total stellar mass). We selected metal rich models (around twice solar metallicity) for all the models, in agreement with previous work  \citep[e.g.][]{Feldmeier-Krause:2017kq,Schultheis:2019aa,Schultheis:2021wf}. To simulate the extinction, we used the extinction curve in \citet{Nogueras-Lara:2020aa}, and the values obtained in Table\,3 of \citet{Nogueras-Lara:2018aa} for the reddening of each of the extinction groups. We applied the extinction randomly assigning extinction values to each star from a Gaussian distribution centred on the mean extinction value for each of the extinction groups. The standard deviation of the Gaussian distributions was $d A_{K_s}\sim0.1$\,mag, in agreement with the extinction distributions in \citet{Nogueras-Lara:2018aa}. We randomly generated the photometric uncertainties assuming a Gaussian distribution for each of the stars with a standard deviation of 0.025\,mag, that correspond to the expected mean uncertainty for the used data at the RC magnitude for $H$ and $K_s$. Figure\,\ref{synthetic} shows the obtained result. The black arrows in the figure depicts the obtained over-densities that are in agreement with the features obtained in the analysis of the target region, and discussed in the previous section.

\subsubsection{Simulation of $K_s$ luminosity functions}

\begin{figure}[htp]
      \begin{center}
   \includegraphics[width=0.9\linewidth]{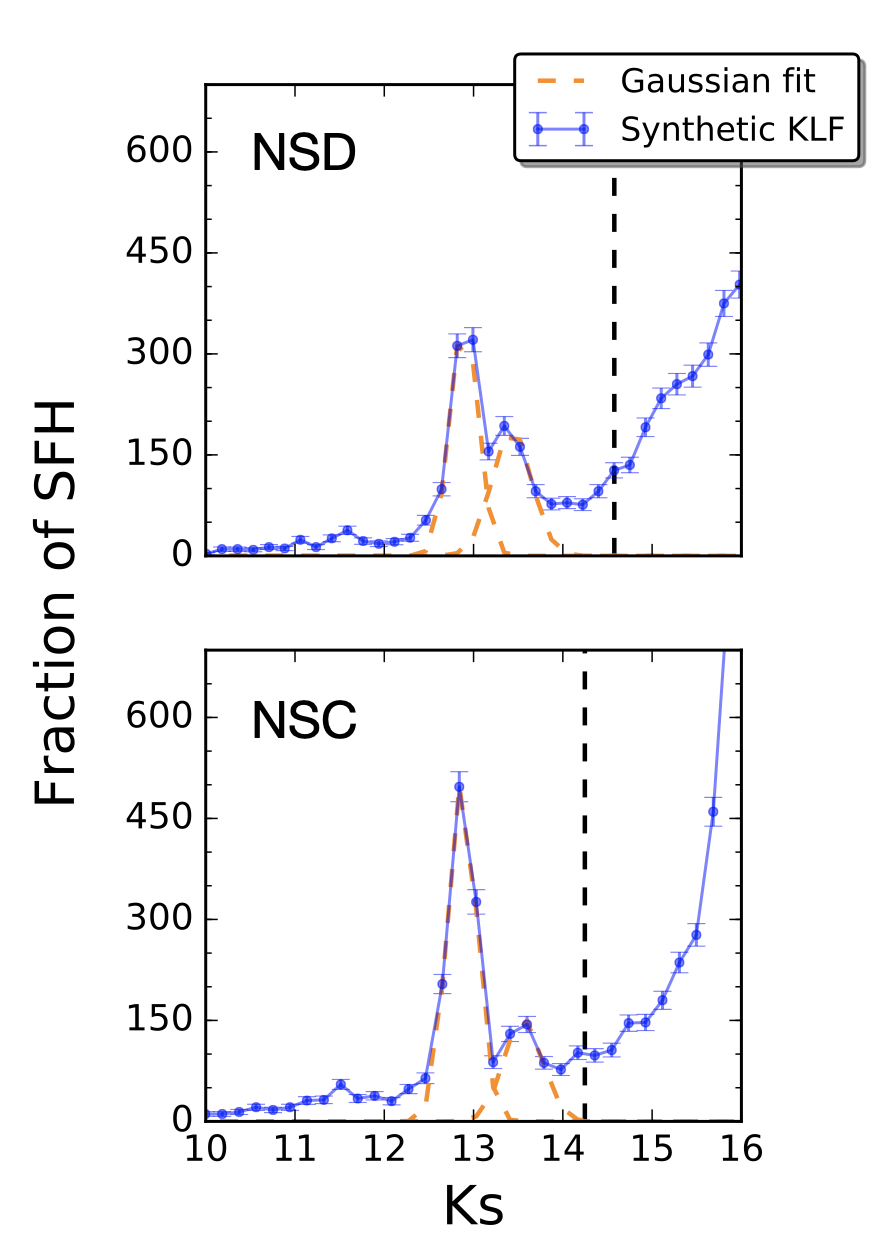}
   \caption{Synthetic de-reddened $K_s$ luminosity functions for the NSD (upper panel) and the NSC (lower panel) corresponding to one of the MC realisations. The black dashed lines indicate the 2/3 completeness limit due to crowding considering de-reddened photometry. The orange dashed lines indicate Gaussian fits to the RC features.}

   \label{KLF_synthetic}
      \end{center}
    \end{figure}

We also built synthetic $K_s$ luminosity functions (KLFs) for each of the stellar populations to compare them with the results obtained in Sect.\,\ref{discu}. We used the corresponding theoretical models, assuming the previously explained uncertainties. To simulate the scatter associated with a real de-reddened KLF \citep[e.g. Sect.\,6 in ][]{Nogueras-Lara:2021wj}, we varied the magnitude of each star assuming a Gaussian distribution centred on its real magnitude with a standard deviation of 0.1\,mag. Figure\,\ref{KLF_synthetic} shows the obtained KLFs for each of the stellar populations. We stress that this is a simple model that aims to check our results considering only the stellar populations that significantly contribute to the RC features, assuming that all the stars are at the same distance. Therefore, the RC features are more prominent in comparison with real data.

We computed the distance between the Gaussian features, the relative fraction of stars belonging to each RC feature, and their associated uncertainties, resorting to 100 Monte Carlo (MC) simulations. For each of them, we produced the KLF generating the corresponding magnitude uncertainties and the scatter associated to the de-reddening process, as previously explained. We fitted the RC features of each KLF using a two-Gaussian model for both cases, the NSD and the NSC. For the NSC, we assumed a single Gaussian to account for the bright RC for simplicity. We obtained mean magnitude differences of $dK_{{s0}\ NSD} = 0.51\pm0.05$\,mag and $dK_{{s0}\ NSC} = 0.66\pm0.03$\,mag, being the distance in magnitude larger for the NSC stellar population, as obtained in Sect.\,\ref{discu}. The uncertainties correspond to the standard deviation of the results obtained for each MC sample. Computing the ratio between the number of stars in each Gaussian feature, we obtained average values of $ratio_{NSD} = 1.25\pm0.28$ and $ratio_{NSC} = 2.72\pm0.29$, where the uncertainties were estimated using the standard deviation of the distributions. We computed mean experimental values from the Gaussian fits in Fig.\,\ref{vertical_bins} averaging over the values obtained for each vertical cut (and combining the bright features for the high extinguished stellar population). We obtained $ratio_{low} = 1.14\pm0.06$ and $ratio_{high} = 2.87\pm0.05$, where the uncertainties correspond to the standard deviation of the measurements. Our results show that the relative fraction of stars between the features is different for each extinction group and that it is compatible with the expected stellar population from the NSD and the NSC. Nevertheless, the computed ratios do not correspond with the real stellar ratios between the RC from different ages and/or the RGBB. This would require a more in-detail analysis taking into account the exponential background associated to the KLF \citep[for further details see e.g.][]{Wegg:2013kx,Nogueras-Lara:2018ab}.

\section{Extinction maps}
\label{emaps}

The detection of the low- and the high-extinguished stellar groups associated to the NSD and the NSC, respectively is possible given the different extinction associated to each of the stellar population. In this way, we also produced extinction maps to compute the extinction variations.

To create the extinction maps associated to each of the group of stars, we used the method described in \citet{Nogueras-Lara:2018aa,Nogueras-Lara:2018ab,Nogueras-Lara:2019ad}. Namely, we used RC and red giant stars fulfilling $K_s\in[13.75, 15.75]$\,mag and $H-K_s\in[1.35, 1.65]$\,mag, for the NSD stellar population, and with $K_s\in[14.5, 16.75]$\,mag and $H-K_s\in[1.8, 2.175]$\,mag, for the NSC. We defined a pixel size of 3 arc-second and computed the associated extinction to each pixel using the five closest reference stars (within a maximum radius of $7.5''$). We computed the extinction using the equation:

\begin{equation}
\label{eq_ratio}
 A_{K_s} = \frac{K_s-H-(K_s-{H})_0}{(A_{K_s}/A_{H})-1}\hspace{0.5cm} ,
\end{equation}
\vspace{0.2cm}

\noindent where $A_{K_s}$ is the extinction in the $K_s$ band, $H$ and $K_s$ are the photometric measurements, the subindex $0$ indicates intrinsic colour, and $A_{K_s}/A_H$ is the used extinction curve and equals to $1.84\pm0.03$ \citep{Nogueras-Lara:2019ac,Nogueras-Lara:2020aa}. To take into account the different distances of the stars to a given extinction map pixel, we used an inverse distance weight method, as explained in Sect.\,7 in \citet{Nogueras-Lara:2018aa}. 

We also built uncertainty maps using a Jackknife resampling method, computing the extinction variation systematically excluding one of the stars used for the extinction calculation associated to each pixel \citep[see Sect. 7 in][]{Nogueras-Lara:2018aa}. The systematic uncertainties were estimated varying the quantities involved in Eq.\,\ref{eq_ratio} within their associated uncertainties.

\begin{figure*}[htp]
   \includegraphics[width=\linewidth]{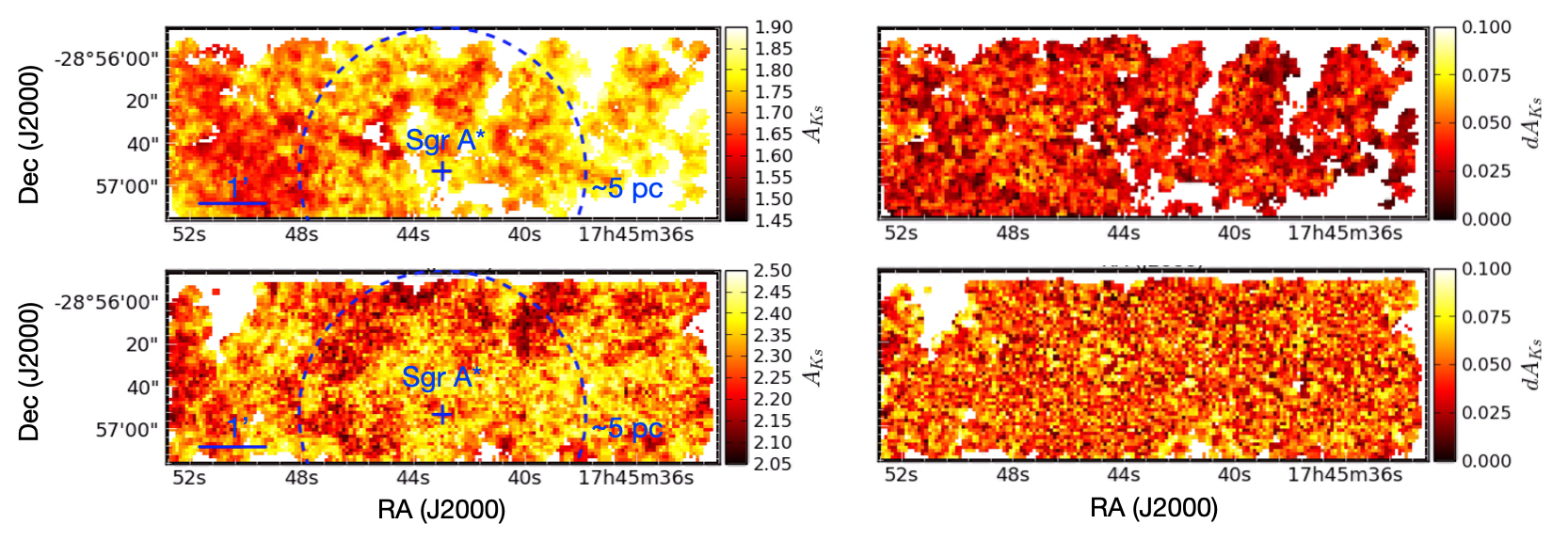}
   \caption{Left column: extinction maps obtained for the low-extinguished group of stars (upper panel), and the high-extinguished one (lower panel). The colour scale is linear and has different limits for each extinction map according to its dynamical range. The blue dashed line indicate the effective radius of the NSC \citep[e.g.][]{gallego-cano2019}. The position of Sgr\,A* is also indicated. Right column: associated uncertainty maps computed as indicated in Sect.\,\ref{emaps}. The colour scales are the same for both uncertainty maps. White pixels in the maps indicate that the number of reference stars were not enough to compute an associated extinction value.}

   \label{ext_maps}
    \end{figure*}

Figure\,\ref{ext_maps} shows the obtained extinction maps and their associated uncertainties. The mean extinction values are $A_{K_s\ NSD}=1.70\pm0.07$\,mag and $A_{K_s\ NSC}=2.27\pm0.09$\,mag. The uncertainties refer to the standard deviation of the value for each map. The mean statistical uncertainty obtained from the uncertainty maps is $\sim2\,\%$, and the systematic uncertainty is $\sim5\,\%$.

The obtained average extinctions are significantly different and point towards two different layers. We also observed that the variation of the extinction across the field associated to the NSC seem to happen on shorter spatial scales in comparison with the extinction map obtained for the NSD (see Fig.\ref{ext_maps}). This is in agreement with the previous work by \citet{Nogueras-Lara:2018aa} using an independent data set \citep[the central pointing of the GALACTICNUCLEUS survey][]{Nogueras-Lara:2019aa}. Therefore, the extinction maps are consistent with having a first extinction layer associated to the interstellar medium in front of the GC (the first extinction map, upper panels in Fig.\ref{ext_maps}), and a secondary layer corresponding to extinction inside the NSD and the CMZ, that varies on shorter spatial scales (arc-seconds), corresponding to the secondary layer (bottom maps in Fig.\ref{ext_maps})). In this way, the low-extinguished stellar group agrees well with the stellar population from the NSD, and the high-extinguished one correspond to the stellar population from the NSC.

\section{NSD contamination of the NSD region in the CMD}

The strong variation of the extinction makes it possible to identify the NSD and the NSC using the CMD $K_s$ vs. $H-K_s$. Nevertheless, there is some correlation between the extinction layers corresponding to each component as shown in Fig.\,\ref{ext_maps}. Therefore, some pollution from the NSD in the NSC feature is expected. To estimate this contamination, we divided the target regions for each of the extinction groups analysed in Fig.\,\ref{CMD} into small vertical cuts of 0.05\,mag and computed the number of stars belonging to each of them. Averaging over the results for each of the groups, we obtained the mean number of stars for each component. Assuming that the density of the NSD stars is approximately constant for the covered extinction range, we can estimate an upper limit to the stars from the NSD present in the NSC sample, computing the ratio between the mean number of stars of each component. We obtained that up to $\sim30$\,\% of the stars in the NSC feature might belong to the NSD. 

We also tested the influence of the completeness on this result. We used the technique presented in Sect.\,3.1.2 in \citet{Nogueras-Lara:2020aa}. We selected a completeness reference level (60\,\%) in the CMD and randomly removed stars in the CMD, in agreement with the completeness solution, to normalise them with respect to the reference level. We created 100 MC samples using this technique and estimated the possible contamination from the NSD on the NSC for each of them, as previously explained. We did not observe any significant difference with respect to the previous result.

\section{The stellar population across different line-of-sights}
\label{different_los}

\begin{figure*}[htp]
   \includegraphics[width=\linewidth]{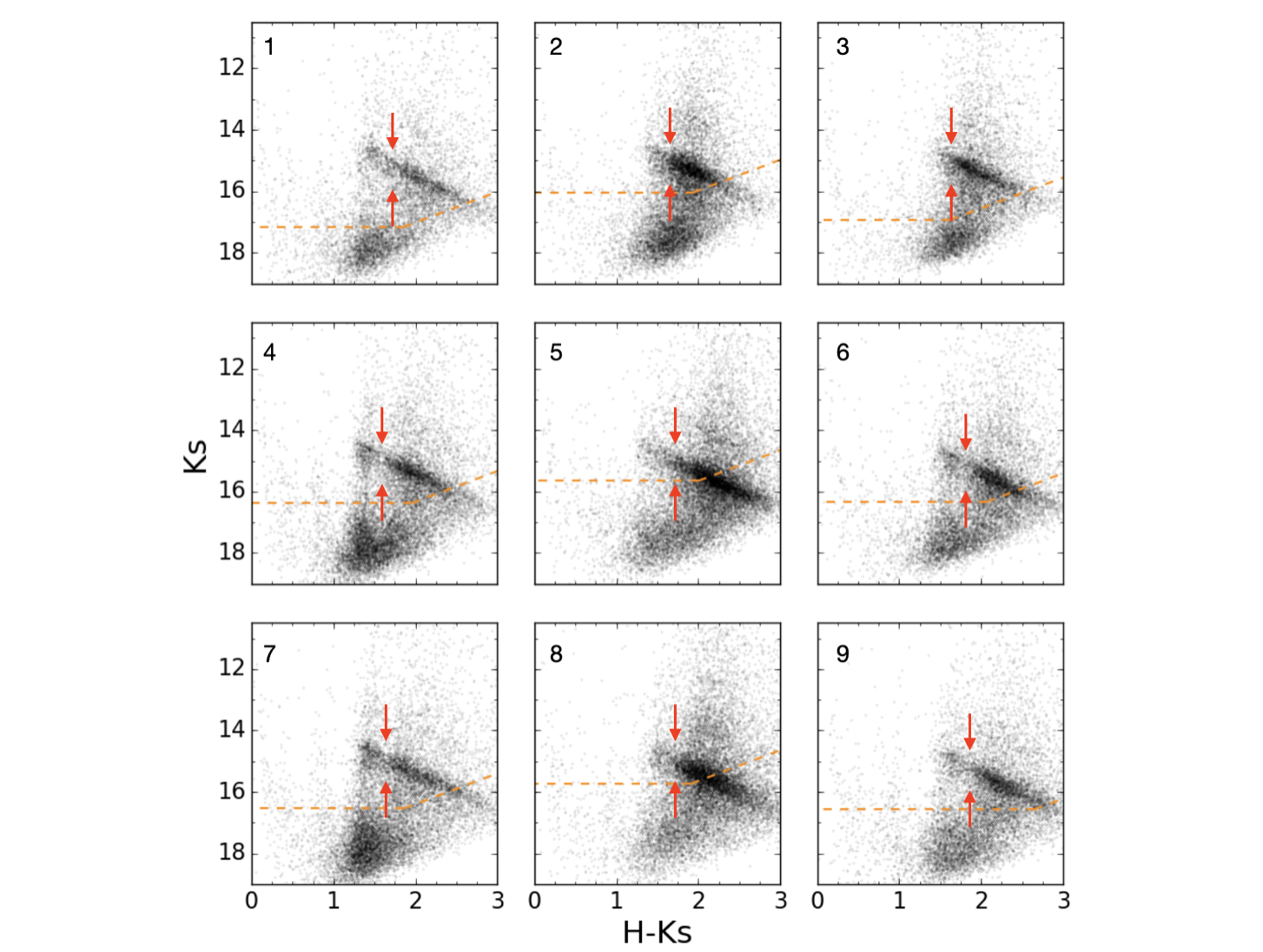}
   \caption{CMDs corresponding to different lines-of-sight across the observed field. The numbers in black show the associated region with respect to Fig.\,\ref{scheme}. The red arrows indicate the cut between the different stellar populations observed. The stellar population corresponding to the NSD is more clearly separated for larger differences in extinction between the stellar groups (in particular it is clearly visible in panels \#4 and \#5). The orange dashed line represents the 2/3 completeness level.}

   \label{CMD_los}
    \end{figure*}

We also analyse the distribution of both extinction groups (corresponding to the NSD and the NSC), in the CMD across different lines-of-sight. For this, we divided the observed field into 9 equally-sized regions, as shown in Fig.\,\ref{scheme}. Figure\,\ref{CMD_los} shows the CMDs associated to each of the regions, where the completeness due to crowding is computed for each of them, following the method explained in Sect.\,\ref{dsp}. We found that the two detected stellar populations appear even more clearly separated for the majority of the regions. This separation correlates well with the difference in extinction between the extinction groups. In this way, we calculated the extinction difference between groups for the different regions using the extinction maps previously computed (see Sect.\,\ref{emaps}). We obtained that the difference is $\Delta A_{K_s}\sim0.6$\,mag for all the regions except for \#2 and \#3, where the difference is $\sim0.1$\,mag smaller and the two groups are less clearly separated.

\subsection{De-reddened $K_s$ luminosity functions}

To assess the presence of different stellar populations between both extinction groups, we de-reddened the RC features for each of them. We chose the stars belonging to each extinction group applying the selection boxes adopted in Sect.\,\ref{dsp} for all the regions. We decided to de-redden the photometry using a star-by-star basis, applying Eq.\,\ref{eq_ratio} for each of the target stars, as explained in \citet{Nogueras-Lara:2021uq}. This method allowed us to have an extinction value for each of the stars in the selection boxes, avoiding the regions without any extinction value appearing in the extinction maps (that would considerably reduce the number of de-reddened stars for some of the regions, biasing the results). 

\begin{figure*}[htp]
   \includegraphics[width=\linewidth]{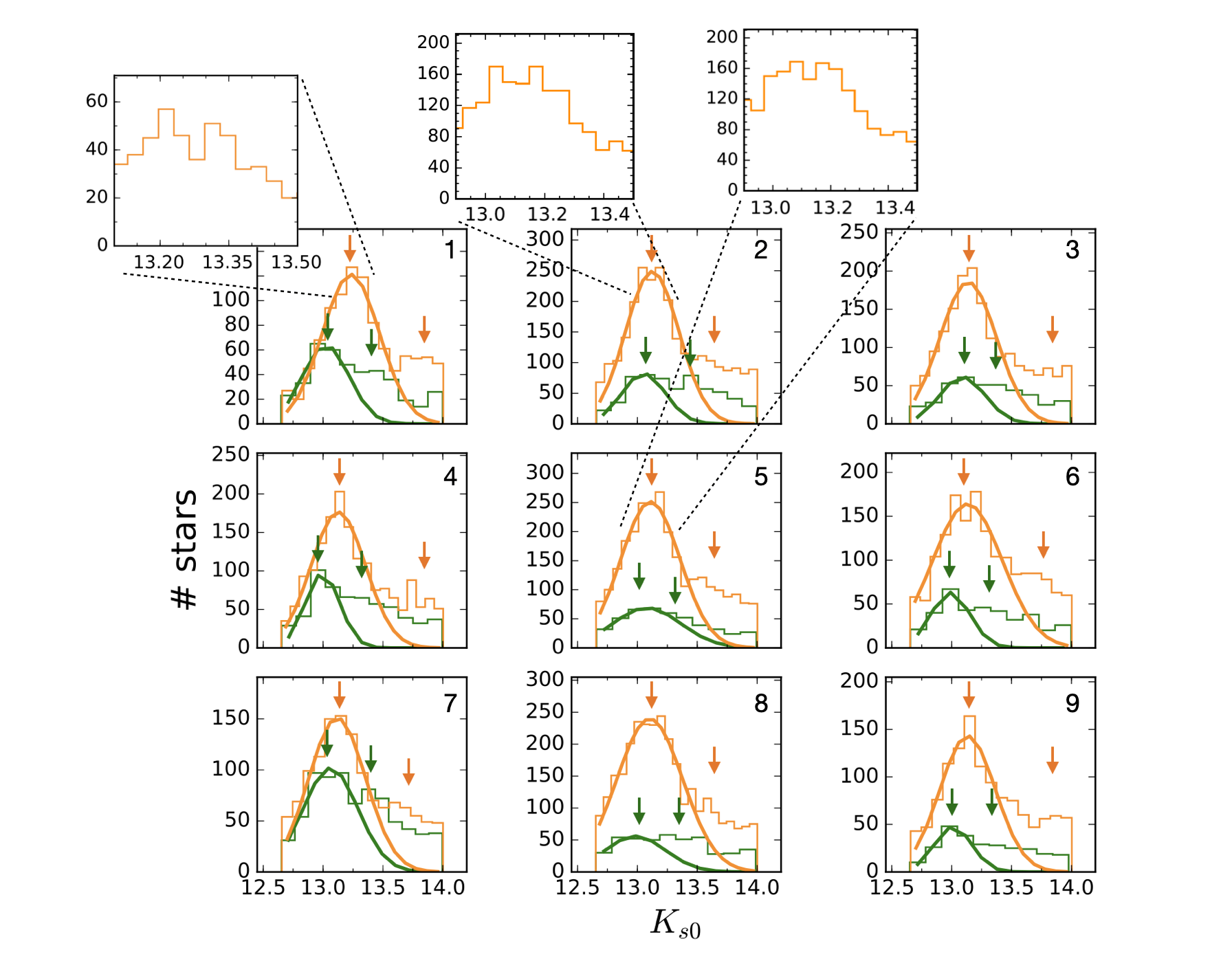}
   \caption{$K_s$ luminosity functions observed for each of the regions in Fig.\,\ref{CMD_los}. The green and orange histograms correspond to the low- and the high-extinguished stellar groups, respectively, following the selection criteria in Sect.\,\ref{dsp}. The green and orange solid lines indicate a Gaussian fit to the brighter peak of the corresponding histogram. The green and orange arrows mark the features observed in the KLFs. The zoom-in panels show how the bright peak of the high-extinguished stellar population corresponds to two close peaks.}

   \label{KLF_regions}
    \end{figure*}

We produced KLFs for each region using the de-reddened $K_s$ values. Figure\,\ref{KLF_regions} shows the obtained results. For each luminosity function, we computed the bin width using the "auto" option implemented in the python function numpy.histogram \citep{Harris:2020aa}. It calculates the maximum of the Freedman-Diaconis \citep{Freedman1981} and the Sturges \citep{doi:10.1080/01621459.1926.10502161} estimators. We observed two main features whose magnitude difference and relative number of stars are significantly different between the KLFs corresponding to each of the different extinction groups, for all the regions. This confirms that there are different stellar populations associated with each of the extinction groups. Nevertheless, the KLFs corresponding to the high-extinguished stellar population appeared to present two peaks instead of the three ones detected in Sect.\,\ref{dsp}. This effect can be partially due to a smoothing of the data associated with the de-reddening process \citep[e.g.][]{Nogueras-Lara:2020aa}. Moreover, there is a systematic effect introduced by the bin width selection, that has an average size of $\sim0.1$\,mag. Therefore, it is not possible to observe a difference of less than $0.2$\,mag. To check this, we selected different smaller bin widths and analysed the KLF region corresponding to the main peak found for the high-extinguished stellar population. The zoom-in panels in Fig.\,\ref{KLF_regions} show the result for three regions as example, where the expected two-peaks structure is clearly visible, in agreement with the result obtained in Sect.\,\ref{dsp}. Therefore, this analysis is consistent with the detection of different stellar populations belonging to the NSD and the NSC, respectively.

\subsection{Distance to the features}

We computed the distance to each of the stellar populations using the mean value of the de-reddened RC features for each region. For this, we fitted a Gaussian to the brightest RC peak detected for each KLF. For the high-extinguished stellar population, the Gaussian fit includes the first and the secondary peaks detected in Sect.\,\ref{dsp}, due to the bin-width size, as it was previously explained. We did not consider the faintest peak for this population given that it does not correspond to the RC, but to the RGBB. Regarding the secondary peak from the low-extinguished stellar population, we did not use it given that it appears to be associated with the $\sim1$\,Gyr star formation event in the NSD \citep{Nogueras-Lara:2019ad}, and the RC brightness varies significantly ($\sim0.5$\,mag) at this age \citep[e.g.][]{Girardi:2016fk,Nogueras-Lara:2019ad}, hampering the use of the distance modulus to estimate the distances.

We computed the distance to each of the fitted features via the distance modulus. We assumed an absolute magnitude for RC stars of $M_{K_s} =-1.59\pm0.04$, as done in \citet{Nogueras-Lara:2021uq}, averaging over the values obtained by \citet{Groenewegen:2008wj}, \citet{Hawkins:2017aa}, and  \citet{Chan:2020ab}. We also applied a population correction factor of $\Delta M_K = -0.07\pm0.07$, as explained in \citet{Schodel:2010fk}. Table\,\ref{dist} shows the obtained results. The uncertainties are systematic, with the statistical ones being negligible. We estimated the systematic uncertainty quadratically propagating the uncertainties involved in Eq.\,\ref{eq_ratio}. We also tested that a more in-detail analysis considering the exponential background associated to the KLF \citep[e.g.][]{Wegg:2013kx,Nogueras-Lara:2018ab} does not influence our results in any significant way within the uncertainties. We concluded that within the uncertainties, the computed distances agree well with the measured distance to the GC of $\sim 8$\,kpc. Therefore, this analysis agrees well with having stellar populations from the NSD and the NSC. We also observed that the distance towards the low-extinguished group appears to be somewhat smaller than for the high-extinguished one. Averaging over the distances obtained for each of the groups, we obtained $d_1=8.16\pm0.14$\,kpc and $d_2=8.52\pm0.15$\,kpc, for the low- and the high-extinguished one, respectively, where the uncertainties were estimated using the standard deviation of the distribution. This is consistent with the observation of the closer edge of the NSD, given its expected radius of $\sim 200$\,pc \citep[e.g.][]{Launhardt:2002nx,Nishiyama:2013uq,gallego-cano2019}. Moreover this is also compatible with the lower extinction measured for the NSD, given that it would correspond to a slightly closer component in comparison with the NSC.

\begin{table}
\caption{Average de-reddened RC magnitudes and distances.}
\label{dist} 
\begin{center}
\def\arraystretch{1.4}
\setlength{\tabcolsep}{3.8pt}
\begin{tabular}{ccccc}
 &  &  &  & \tabularnewline
\hline 
\hline 
Region & $K_{s0\ 1}$ & $K_{s0\ 2}$ & $d_1$ & $d_2$\tabularnewline
 & (mag) & (mag) & (kpc) & (kpc)\tabularnewline
\hline 
1 & 13.02 & 13.23 & 8.1 $\pm$ 0.5 & 8.9 $\pm$ 0.5\tabularnewline
2 & 13.05 & 13.12 & 8.2 $\pm$ 0.5 & 8.5 $\pm$ 0.5\tabularnewline
3 & 13.09 & 13.14 & 8.4 $\pm$ 0.5 & 8.5 $\pm$ 0.5\tabularnewline
4 & 13.00 & 13.13 & 8.0 $\pm$ 0.4 & 8.5 $\pm$ 0.5\tabularnewline
5 & 13.08 & 13.11 & 8.3 $\pm$ 0.5 & 8.4 $\pm$ 0.5\tabularnewline
6 & 12.99 & 13.13 & 8.0 $\pm$ 0.4 & 8.5 $\pm$ 0.5\tabularnewline
7 & 13.06 & 13.12 & 8.3 $\pm$ 0.5 & 8.5 $\pm$ 0.5\tabularnewline
8 & 12.98 & 13.10 & 8.0 $\pm$ 0.4 & 8.4 $\pm$ 0.5\tabularnewline
9 & 13.01 & 13.13 & 8.1 $\pm$ 0.4 & 8.5 $\pm$ 0.5\tabularnewline
\hline 
 &  &  &  & \tabularnewline
\end{tabular}

\end{center}

\textbf{Notes.} $K_{s0\ 1}$ and $K_{s0\ 2}$ indicate the de-reddened $K_s$ magnitudes corresponding to the main peaks in the KLFs for the low- and the high-extinguished stellar groups, respectively. The associated uncertainties are $\Delta K_s \sim0.12$\,mag and correspond to systematics, being the statistical uncertainty negligible. $d_1$ and $d_2$ correspond to the associated distance of each of the stellar groups. For relative comparison between the distances corresponding to the different features, the uncertainty is $\sim20\,\%$ lower. 
 \end{table}

\section{Conclusions}

In this paper, we analysed the $H$ and $K_s$ photometry from a region of $8.2'\times 2.8'$ centred on the NSC. We built a CMD $K_s$ vs. $H-K_s$, and detected the presence of two different stellar groups in the RC associated with different extinctions. The different RC features detected for each of the extinction groups, and the magnitude distance between them, agree well with the stellar populations and SFHs expected for the NSD and the NSC \citep{Nogueras-Lara:2019ad,Schodel:2020aa}. We checked  this hypothesis using a control field in the NSD and a synthetic CMD reproducing the stellar populations from the NSD and the NSC. Using simulations, we concluded that the different RC features can be explained considering an intermediate-age stellar population in the NSC that is not present in the NSD \citep{Schodel:2020aa}. Analogously, the faint RC feature measured for the NSD does not have a counterpart in the NSC and corresponds to a star formation event around 1\,Gyr ago \citep[e.g.][]{Rui:2019aa,Nogueras-Lara:2019ad,Nogueras-Lara:2021uq}.

The presence of these different stellar groups with different extinctions were previously reported by \citet{Nogueras-Lara:2018aa}, where the low extinction group was tentatively explained as a stellar feature in front of the central molecular zone \citep[e.g.][]{Morris:1996vn}. According to our results, we believe that this scenario is improbable. Building extinction maps, we conclude that the low-extinguished stellar population belongs to the NSD.

Finally, we computed the distance towards the NSD and the NSC creating de-reddened KLFs and fitting them with a Gaussian model. The results agree with the expected GC distance. Moreover, we found some evidence of a slightly closer distance to the stellar population from the NSD, that appears to be $360\pm200$\,pc closer to us. This is in agreement with the observation of the stars from the closest edge of the NSD, considering its radius of $\sim 200$\,pc.

Our results show that the NSD and the NSC are detectable due to significantly different extinctions along the line-of-sight towards the NSC, and point towards the presence of different stellar populations associated with a different SFH and formation scenario of both components.

\begin{acknowledgments}
      {\bf Acknowledgments}. This work is based on observations made with ESO
      Telescopes at the La Silla Paranal Observatory under programmes
      IDs 195.B-0283 and 091.B-0418. We thank the staff of
      ESO for their great efforts and helpfulness. F.N.-L. acknowledges the sponsorship provided by the Federal Ministry for Education and Research of Germany through the Alexander von Humboldt Foundation. F.N.-L. and N.N. gratefully acknowledge support by the Deutsche Forschungsgemeinschaft (DFG, German Research Foundation) – Project-ID 138713538 – SFB 881 ("The Milky Way System", subproject B8). R.S. acknowledges financial support from the State
Agency for Research of the Spanish MCIU through the "Center of Excellence Severo
Ochoa" award for the Instituto de Astrof\'isica de Andaluc\'ia (SEV-2017-0709). R.S.  acknowledges financial support from national project
PGC2018-095049-B-C21 (MCIU/AEI/FEDER, UE).
\end{acknowledgments}

%% For this sample we use BibTeX plus aasjournals.bst to generate the
%% the bibliography. The sample631.bib file was populated from ADS. To
%% get the citations to show in the compiled file do the following:
%%
%% pdflatex sample631.tex
%% bibtext sample631
%% pdflatex sample631.tex
%% pdflatex sample631.tex

\bibliography{../../../../BibGC.bib}{}
\bibliographystyle{aasjournal}

%% This command is needed to show the entire author+affiliation list when
%% the collaboration and author truncation commands are used.  It has to
%% go at the end of the manuscript.
%\allauthors

%% Include this line if you are using the \added, \replaced, \deleted
%% commands to see a summary list of all changes at the end of the article.
%\listofchanges

\end{document}